\documentclass[apj]{emulateapj}

\newcommand{\afe}{[$\alpha$/Fe]}
\newcommand{\zh}{[Z/H]}

\slugcomment{Accepted for publication in ApJ; May 8th, 2006}

\shorttitle{Extremely $\alpha$-enriched Globular Clusters in Early-Type Galaxies} 
\shortauthors{Puzia, Kissler-Patig, \& Goudfrooij}

\begin{document}

\title{Extremely $\alpha$-enriched Globular Clusters in Early-Type Galaxies:\\A Step towards the Dawn of Stellar Populations?}

\author{Thomas H. Puzia\altaffilmark{1,2}, Markus
  Kissler-Patig\altaffilmark{3}, \& Paul Goudfrooij\altaffilmark{1}} 

\altaffiltext{1}{Space Telescope Science Institute, 3700 San Martin Drive,
    Baltimore, MD 21218, USA.}
\altaffiltext{2}{ESA Research Fellow, Space Telescope Division of ESA.}
\altaffiltext{3}{European Southern Observatory, Karl-Schwarzschild-Str.\ 2,
    85748 Garching bei M\"unchen, Germany.}

\begin{abstract}
We compare \afe, metallicity, and age distributions of globular clusters
in elliptical, lenticular, and spiral galaxies, which we derive from Lick
line index measurements.~We find a large number of globular clusters in
elliptical galaxies that reach significantly higher \afe\ values
(\afe~$\!>\!0.5$) than any clusters in lenticular and spiral
galaxies.~Most of these extremely $\alpha$-enriched globular clusters are
old ($t\!>\!8$ Gyr) and cover the metallicity range $-1\!\la$ \zh\
$\la\!0$.~A comparison with supernova yield models suggests that the
progenitor gas clouds of these globular clusters must have been
predominantly enriched by massive stars ($\ga\!20\, M_{\odot}$) with
little contribution from lower-mass stars.~The measured \afe\ ratios are
also consistent with yields of very massive pair-instability supernovae
($\sim\!130\!-\!190\, M_{\odot}$).~Both scenarios imply that the chemical
enrichment of the progenitor gas was completed on extremely short
timescales of the order of a few Myr.~Given the lower \afe\ average ratios
of the diffuse stellar population in early-type galaxies, our results
suggest that these extremely $\alpha$-enhanced globular clusters could be
members of the very first generation of star clusters formed, and that
their formation epochs would predate the formation of the majority of
stars in giant early-type galaxies.
\end{abstract}


\section{Introduction}
The field of globular cluster system research has contributed countless
important insights to astronomy \citep[e.g.][and references
therein]{shapley18, searle78, zinn85, harris91, az98}.~Because they
survive a Hubble time, globular cluster systems (GCSs) provide important
information on the assembly history of their parent galaxy and the
physical conditions in the star formation episodes during which they
formed.~With the advent of the {\it Hubble Space Telescope}, extragalactic
GCSs in a variety of host galaxies outside the Local Group became
accessible to accurate photometric studies \citep[e.g.][]{whitmore93,
whitmore95, gebhardt99, puzia99, kundu01a, kundu01b,
cote04}.~In the last few years, progress in GCS research experienced a
boost through the availability of efficient multi-object spectrographs,
which facilitate spectroscopy of large samples of extragalactic globular
clusters out to large distances \citep[$\sim\!20$ Mpc; see e.g.][]{kp98,
cohen98, puzia04}.~Spectroscopy provides the best constraints on
the ages and chemical compositions of globular clusters.

The chemical composition of individual extragalactic globular clusters
provides insight on the enrichment histories and timescales of their
parent massive gas clouds.~Because of the different progenitor lifetimes
of type II and type Ia supernovae (SNe) \citep[e.g.][]{tornambe86} and the
different chemical composition of their ejecta \citep[e.g.][]{nomoto97II},
the progenitor gas clouds will undergo different chemical evolution
depending on whether they were enriched on short timescales predominantly
by type II SNe, or on extended timescales by both type II {\it and} Ia
SNe.~Ages and chemical compositions of globular clusters can be used to
reconstruct the early assembly histories of globular cluster systems and
their host galaxies, since since self-enrichment is excluded for all but
the most massive globular clusters \citep{recci05}.~Together with accurate
ages and metallicities, the chemical composition of globular clusters can
be used to reconstruct the early assembly histories of globular cluster
systems.

In this paper, we investigate \afe, metallicity, and age distributions of
individual globular clusters systems as a function of environment (in
elliptical, lenticular and spiral galaxies) based on high-quality
spectroscopy data from the literature. We identify and discuss extremely
$\alpha$-enriched clusters that are (almost) exclusively found in
ellipticals.

\section{Data}
The majority of our analysis is based on the spectroscopic dataset of
\cite{puzia04}, who obtained high-S/N spectra for globular clusters in
seven early-type galaxies (NGC 1380, 2434, 3115, 3379, 3585, 5846, and
7192) with the FORS multi-object spectrograph at ESO's {\it Very Large
Telescope} (VLT).~We augment this sample with high-quality index
measurements from the literature, which are explained in the last section
in \cite{puzia04}.~We refer the reader to this paper for further
details.~In summary, we add to the sample of globular clusters in
elliptical galaxies data from NGC~3610 \citep{strader04} and NGC~4365
\citep{larsen03, brodie05}.~To the sample of globular clusters in
lenticular galaxies we add spectra from NGC~3115 \citep{kuntschner02},
NGC~4594 \citep{larsen02}, and NGC~5128 \citep{peng04}.~The sample of
globular clusters in spiral galaxies is assembled of clusters in the Milky
Way \citep{puzia02b}, M31 \citep{trager98, beasley04, puzia05b}, M81
\citep{schroder02}, and the Sculptor spirals NGC~55, 247, 253, and 300
\citep{olsen04}.

All index measurements were uniformly performed with the passband
definitions of \cite{worthey94} and \cite{worthey97} and are calibrated on
the Lick standard system using Lick index standards.~In the following
analysis we consider only globular clusters which have high-quality
measurements in the indices H$\beta$, H$\gamma_{\rm A}$, H$\delta_{\rm
A}$, $\langle$Fe$\rangle$, Mg$b$, and Mg$_2$ since these enter our age,
metallicity, \afe\ fitting routine.~To guarantee a robust comparison of
the three samples, we select data with measurement uncertainties
$\Delta$H$\beta\leq0.5$, $\Delta$H$\gamma_{\rm A}\leq0.7$,
$\Delta$H$\delta_{\rm A}\leq0.7$, and $\Delta$[MgFe]\arcmin$\leq0.4$ \AA\
\citep[see][]{puzia05}.~The final dataset contains 87, 47, and 59 globular
clusters with high-quality index measurements in elliptical, lenticular,
and spiral galaxies, respectively, and samples the brightest $\sim\!10$\%
of the globular cluster luminosity function.~In terms of calibration
quality and signal-to-noise, it is, to our knowledge, the best-quality
spectroscopic dataset of extragalactic globular clusters currently
available.

\section{Analysis}
\begin{figure}[!ht]
\centering 
\includegraphics[width=6.3cm]{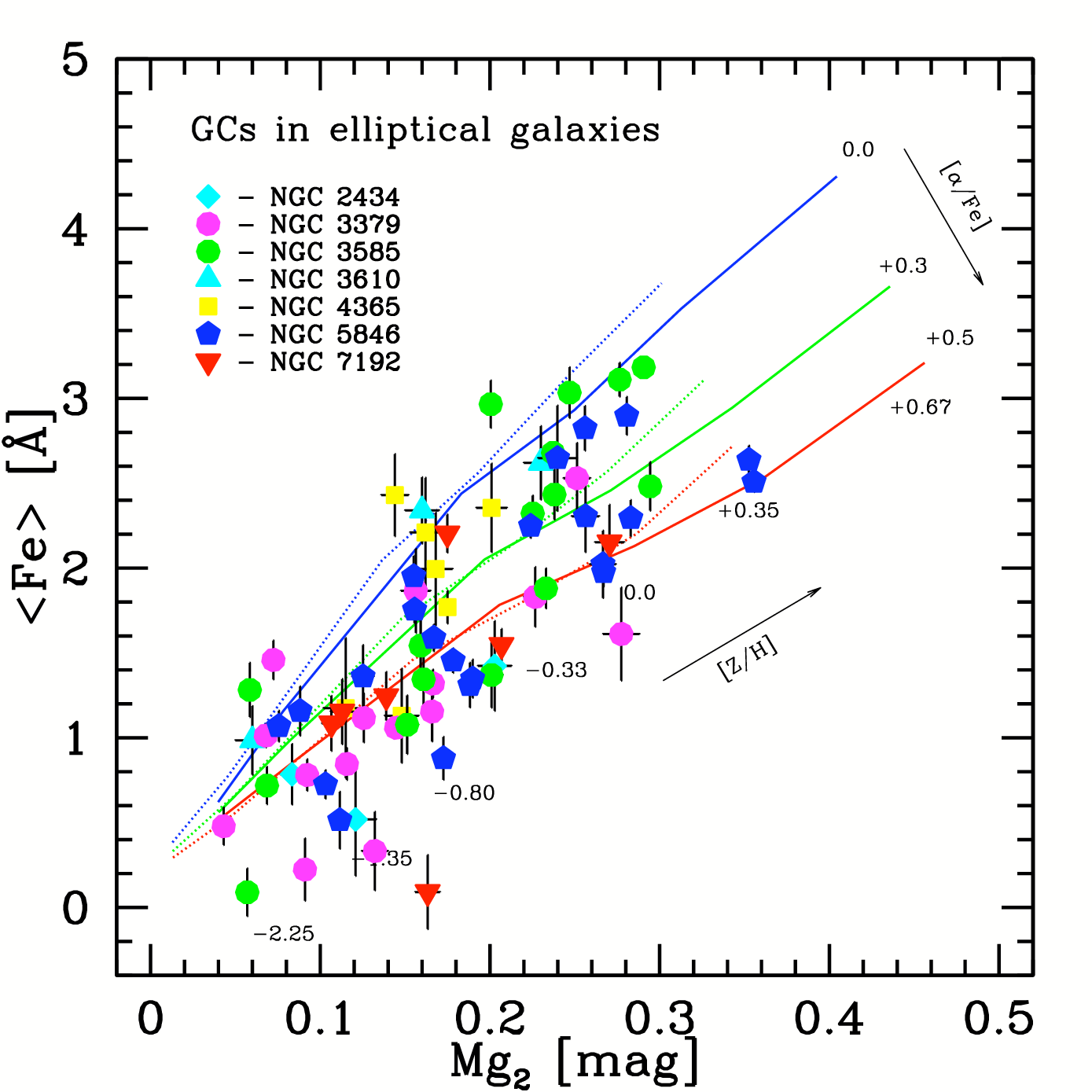}
\includegraphics[width=6.3cm]{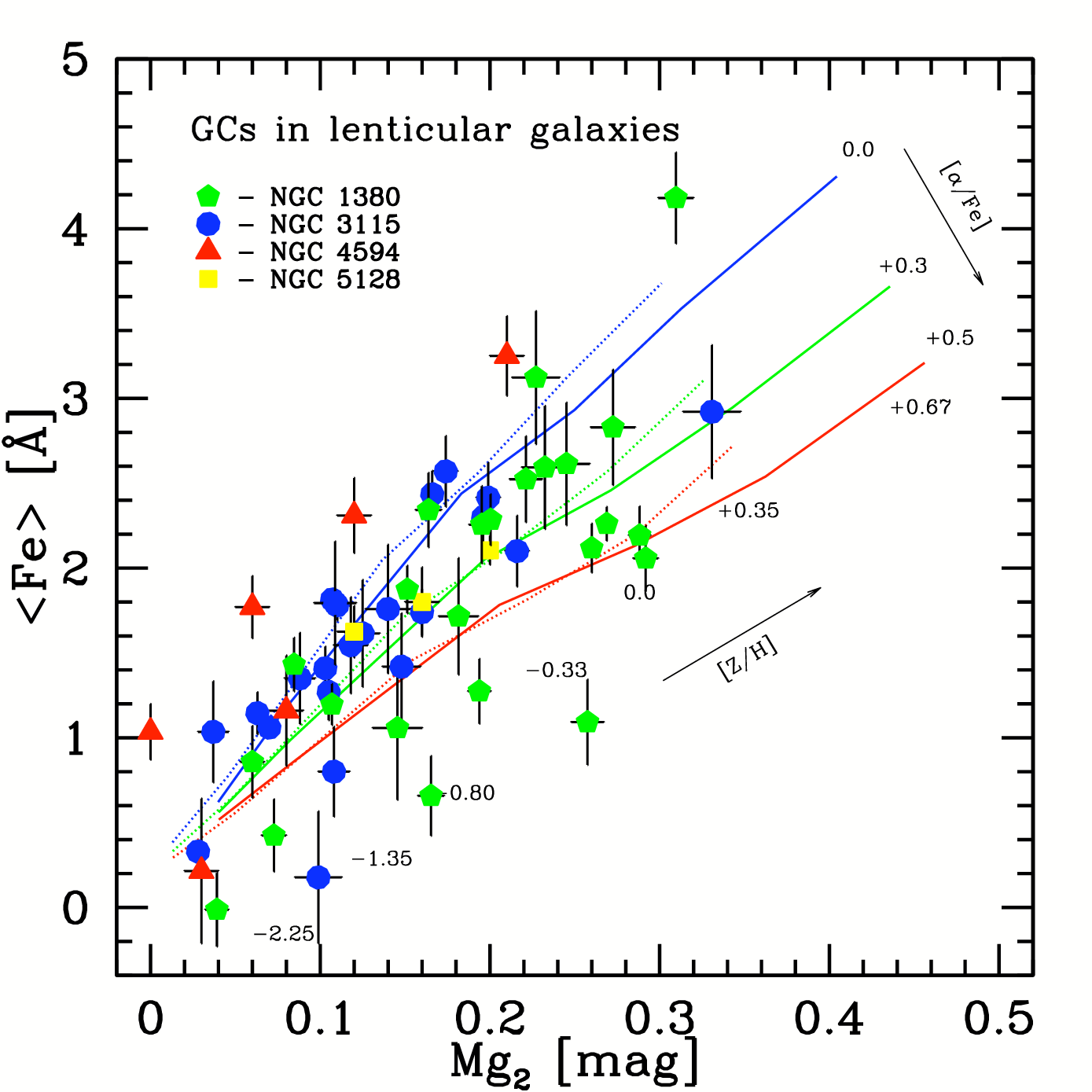}
\includegraphics[width=6.3cm]{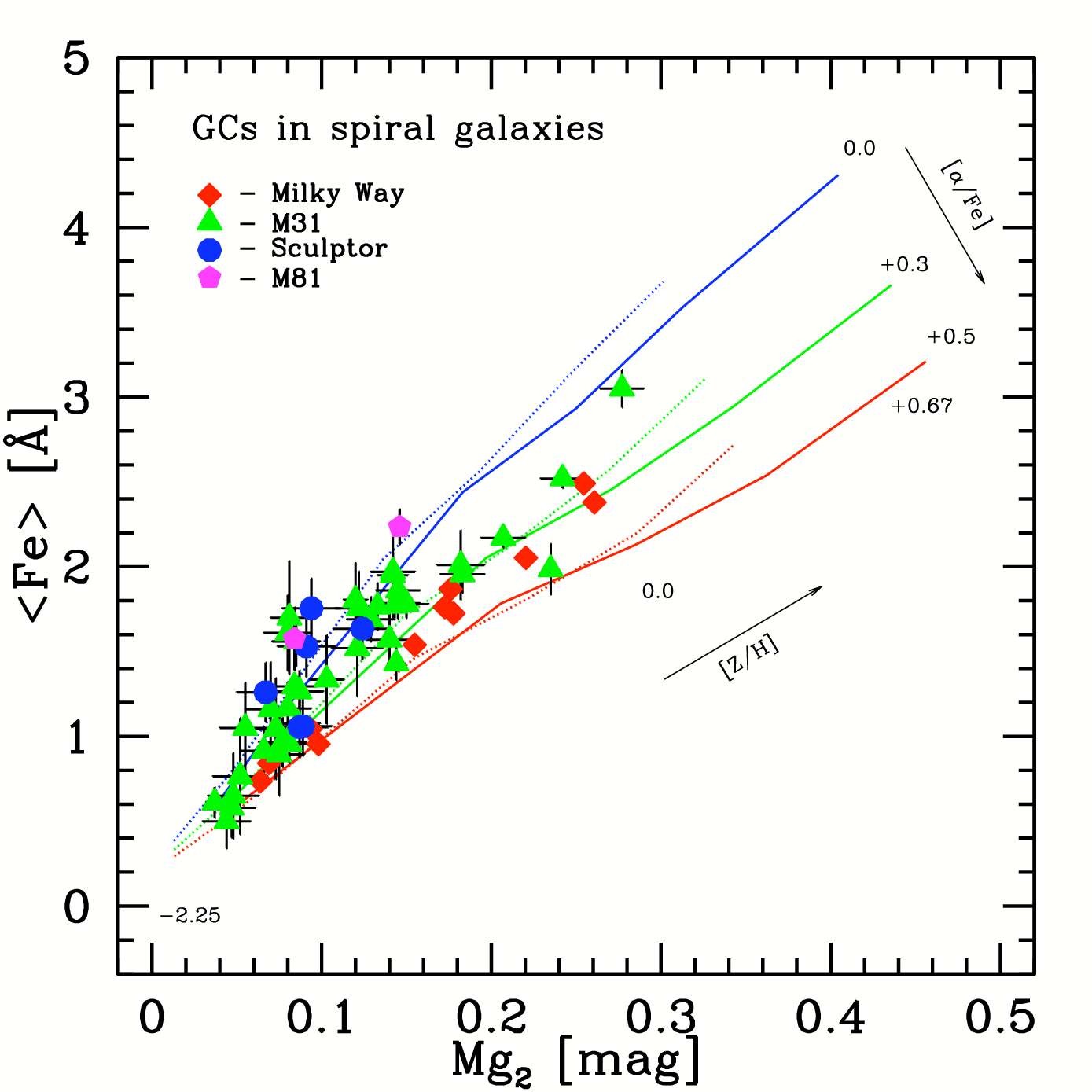}
\caption{This figure shows \afe\ diagnostic plots for globular clusters in
spiral ({\it upper panel}), lenticular ({\it middle panel}), and elliptical
galaxies ({\it bottom panel}).~Different symbols show different globular
clusters in individual host galaxies which are described in the legend of
each panel.~Solid lines show iso-\afe\ tracks for 0.0, 0.3, and 0.5 dex
for a 13 Gyr old stellar population with metallicities from [Z/H]~$=-2.25$
to $+0.67$.~To illustrate the systematics due to age, we show the same
tracks for a 3 Gyr stellar population, indicated by dotted lines.~Note
that the models tracks are virtually independent of age up to solar 
metallicities.~The model predictions were taken from 
\cite{thomas03, thomas04}.}
\label{ps:dgpl}
\end{figure}
We use the technique described in \cite{puzia05} to derive ages,
metallicities, and \afe\ ratios for all sample globular clusters. In
summary, we compare the age-sensitive Balmer line indices H$\beta$,
H$\gamma_{\rm A}$, H$\delta_{\rm A}$ and the metallicity-sensitive indices
$\langle$Fe$\rangle$, Mg$_{2}$ and [MgFe]\arcmin\ to the population
synthesis models of \cite{thomas03} and \cite{thomas04} which provide
predictions for stellar populations with well-defined abundance patterns,
in particular with varying \afe\ ratios.~These models were previously
calibrated on Galactic globular clusters \citep{puzia02b, maraston03} for
which accurate abundance ratios, including Fe-peak and $\alpha$-elements,
were known from high-resolution spectroscopy.~Ages, metallicities, and
\afe\ ratios are derived in an iterative $\chi^{2}$-minimization approach
of diagnostic grids taking into account index measurement errors and
systematic uncertainties of the Lick index system.~Since the error
distributions are similar for all indices that enter the analysis, a
comparison of all three globular cluster samples with the same SSP model
predictions guarantees a robust differential analysis.~Absolute globular
cluster ages have to be interpreted cautiously, since the absolute age
scale is prone to systematic uncertainties in the model calibration and an
unknown contribution from hot horizontal branch stars.~However, our
analysis allows the distinction between old ($t\!>\!8$ Gyr),
intermediate-age ($2\!<\!t\!<\!8$ Gyr), and young ($t\!<\!2$ Gyr) globular
clusters.~The internal metallicity and \afe\ calibrations
are more robust and have systematic uncertainties $\la0.15$ dex.~It is
important to realize that these internal systematics do not affect our
{\it differential} analysis.

\begin{figure}[!t]
\centering 
\includegraphics[bb=24 150 592 718,width=8.5cm]{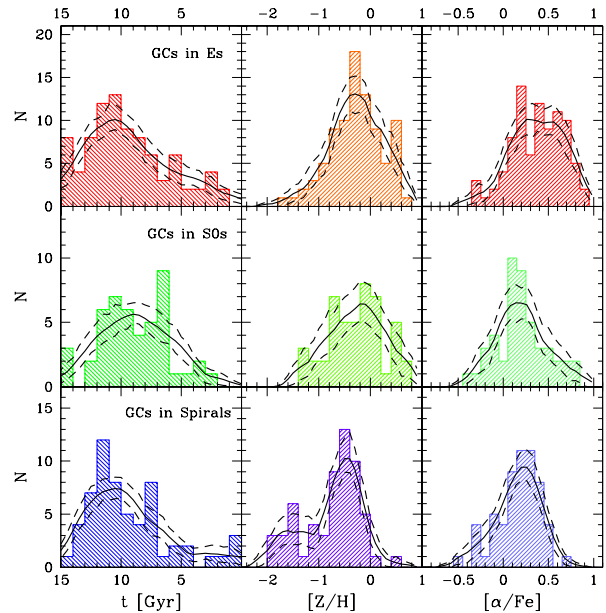}
 \caption{Age, metallicity, and \afe\ distributions of globular clusters in
elliptical ({\it upper row}), lenticular ({\it middle row}), and spiral
galaxies ({\it bottom row}). Solid curves are non-parametric probability
density estimates with their 90\% confidence limits indicated by dashed 
lines.}
\label{ps:amah}
\end{figure}

We present \afe\ diagnostic plots for globular clusters in spiral,
lenticular, and elliptical galaxies in Figure~\ref{ps:dgpl}.
Figure~\ref{ps:amah} shows the derived distributions of globular cluster
ages, metallicities, and \afe\ ratios in elliptical, lenticular, and
spiral galaxies.~The most unbiased representation of these data is
realized via a non-parametric probability density estimate using an
Epanechnikov-kernel \citep[see][for details]{silverman86}, which are
indicated by solid lines.~Dashed lines show bootstrapped 90\% confidence
limits.~The median ages for the elliptical, lenticular, and spiral samples
are $10.2\pm2.1$, $9.4\pm2.2$, and $10.5\pm2.0$ Gyr\footnote{The errors
give the semi-interquartile ranges.}.~The sampling of all three
sub-samples is somewhat biased towards the central regions of galaxies and
therefore towards more metal-rich globular clusters.~In fact, in the
elliptical and lenticular sample there are only few clusters more
metal-poor than [Z/H]~$\approx-1.5$, which correspond to the metal-poor
sub-population of Galactic halo globular clusters
\citep[e.g.][]{zinn85}.~For both the elliptical and lenticular sub-sample
we find a slightly sub-solar mean metallicity, while the spiral sample is
much less affected by this bias and probes significantly lower globular
cluster metallicities down to [Z/H]~$\approx-2.0$.~The corresponding
median metallicities are [Z/H]~$=-0.32, -0.23,$ and $-0.58$ dex, for the
elliptical, lenticular, and spiral sample, respectively.

The perhaps most striking feature in Figure~\ref{ps:amah} is the very
broad and possibly bimodal \afe\ distribution for globular clusters in
elliptical galaxies (upper right panel), with a high-\afe\ ''shoulder''
reaching values up to $\sim\!1.0$ dex.~This shoulder is significantly less
developed in the sample of lenticular galaxies and entirely absent for the
sample of spiral galaxies. Both spiral and lenticular samples have a
median value of \afe\ is $\sim$\,0.2 dex, twice as low as for the sample
of ellipticals.~We emphasize that, although NGC~1380 GCS may slightly
dominate the high-\afe\ values in the lenticular sample, no single
globular cluster system dominates any of the distributions.~Instead,
globular clusters from all galaxies populate the derived distributions rather
homogeneously (see Fig.~\ref{ps:dgpl}).

Although there is some evidence for multiple components in the probability
density estimate of the \afe\ distribution of globular clusters in
ellipticals, the KMM algorithm \citep{ashman94} rejects the hypothesis
that the distribution is different from bimodal at a $\sim\!2\sigma$
level.~The test does not strongly support bimodality, but does provide
additional evidence that the \afe\ distribution for globular clusters in
ellipticals is significantly broader and reaches much higher \afe\ values
than those for globular clusters in lenticular and spiral galaxies.
\section{Discussion}
\begin{figure*}[!t]
\centering 
\includegraphics[bb=13 21 398 398, width=13cm]{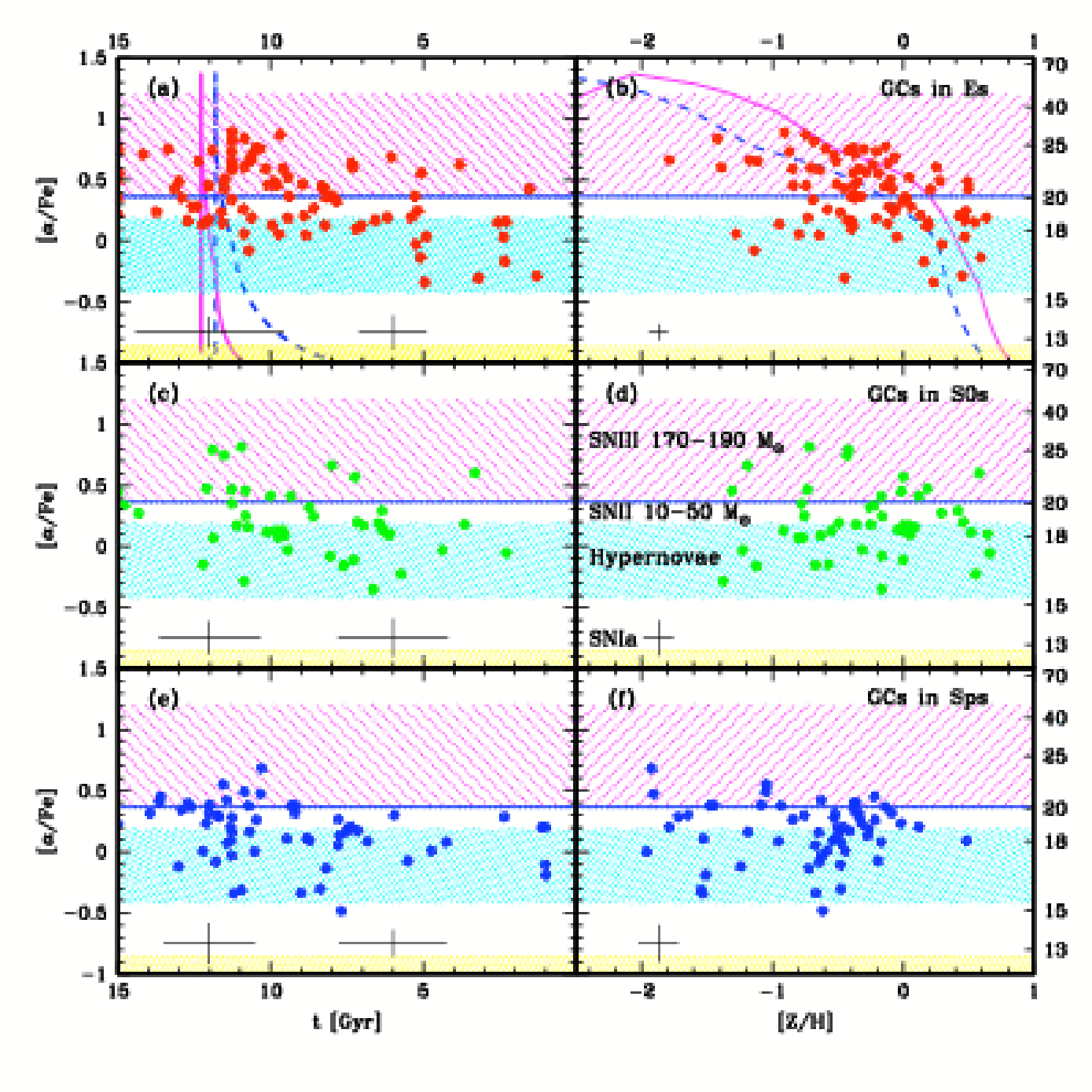}
 \caption{Correlation plots of \afe\ ratios with ages ({\it left panels}) 
and metallicities ({\it right panels}) for globular
clusters in elliptical ({\it upper row}), lenticular ({\it middle row}),
and spiral galaxies ({\it bottom row}).~To illustrate the enrichment by
supernovae of different types, mean \afe\ ratios of the ejecta of type Ia
SNe ({\it yellow shading}), cumulative type II SNe
ejecta with progenitor masses in the range $10\!-\!50 M_{\odot}$ ({\it
thick horizontal line}) integrated over a Salpeter IMF,
hypernovae with very large explosion energies $\ga10^{52}$ ergs ({\it cyan
shading}), and type III pair-instability supernovae
with progenitor masses $170\!-\!190 M_{\odot}$ ({\it magenta hatched
region}) were overplotted (see Sect.~\ref{sec:SNenrich} for details).~The
right ordinate indicates progenitor masses of type II SNe that produce 
the [$\alpha$/Fe] indicated by the left ordinate.~In
the upper panels ($a$ and $b$), we plot two enrichment models for an
elliptical galaxy with $10^{10} M_{\odot}$ ({\it dashed line}) and $10^{12}
M_{\odot}$ initial baryonic mass ({\it solid line}), taken from
\cite{pipino04}.~Small crosses indicate average statistical
errors.~Because of the changing $\Delta t/t$, the statistical age
uncertainties were calculated for the young and old sub-samples split at 8
Gyr.}
\label{ps:amac}
\end{figure*}

While the average properties of the samples were already addressed in
\cite{puzia05}, we focus here on the globular clusters showing extreme
\afe\ values. Figure~\ref{ps:amac} shows \afe\ plotted against age (left
panels) and metallicity (right panels) for globular clusters in
elliptical, lenticular, and spiral galaxies (from top to bottom).~The
direct comparison of GCSs shows important differences:~globular clusters
tend to have higher \afe\ ratios and reach higher metallicities in hosts
with earlier morphological type.~Most of the extremely $\alpha$-enhanced
clusters (\afe~$>0.5$) are old ($t\!\ga\!8$ Gyr), and are mainly found in
elliptical galaxies.~Very few counterparts are found in lenticular and
spiral galaxies.~The metallicities of these extreme \afe\ clusters
span a wide range from metal-poor [Z/H]~$\approx-2$ dex to solar,
concentrating around values between $-1\la$~[Z/H]~$\la0$ dex in
ellipticals and lenticulars, while the few candidates in spirals have
metallicities [Z/H]~$\la-1$ dex.~There is tentative evidence for an
age-\afe\ correlation for globular clusters in elliptical galaxies
(correlation coefficient 0.5; see panel a in Fig.~\ref{ps:amac}), in the
sense that younger globular clusters have on average lower \afe\
ratios.~For metal-rich globular clusters ([Z/H]~$\!>\!-1.0$) in elliptical
galaxies, we find an anti-correlation of \afe\ and metallicity, which
becomes weaker toward later-type hosts.~At face value, these results
suggest that a first generation of globular clusters in elliptical
galaxies was enriched to much higher \afe\ values (i.e.~faster) than in
lenticular and spiral galaxies.
\subsection{Chemical enrichment with different SN types}
\label{sec:SNenrich}

We explore in the following whether the \afe\ ratios of the extremely
$\alpha$-enriched clusters can be reproduced by state-of-the-art supernova
yield models.~For this purpose we compute \afe\ ratios of the ejecta of
type II SNe in the mass range $13\!-\!70\, M_{\odot}$ \citep{nomoto97II},
type Ia SNe \citep{nomoto97Ia}, hypernovae with very large explosion
energies $\ga\!10^{52}$ erg \citep{nakamura01}, and pair-instability SNe
with progenitor masses $>\!160\, M_{\odot}$ \citep{heger02}.~To stay
consistent with the definitions of the SSP models, we compute \afe\ ratios
of the SN yields as the mass of the $\alpha$-elements N, O, Mg, Ca, Na,
Ne, S, Si, and Ti relative to the mass of the Fe-peak elements Cr, Mn, Fe,
Co, Ni, Cu, and Zn.~The results are normalized to the solar abundances
given in \cite{anders89}.~We note that other combinations of
$\alpha$-elements, in particular the addition of carbon, does not affect
the results.

The \afe\ ratio of the cumulative ejecta of core-collapse SNe with
progenitor masses $10\!-\!50\, M_{\odot}$ integrated over a Salpeter IMF is
indicated in Figure~\ref{ps:amac} as a solid horizontal line at
$\sim\!0.36$ dex.~Such an IMF extending to high masses can reproduce the
[$\alpha$/Fe] ratios of virtually all globular clusters in spirals and the
majority of globular clusters in lenticular galaxies, but not the most
extreme \afe\ ratios of globular clusters in elliptical galaxies.~A
significant contribution of massive stars $\ga\!20\, M_{\odot}$ is necessary
to boost the \afe\ beyond $\sim\!0.4$ dex (see Fig.~\ref{ps:amac}).~The
ejecta of pair-instability SNe, which require very metal-poor progenitor
stars, are expected to be rich in $\alpha$-elements \citep{heger02}. Our
calculations indicate that the parent gas clouds of globular clusters can
be enriched to extreme \afe\ ratios by pair-instability SNe with masses
$\sim130\!-\!190\, M_{\odot}$.~We find that hypernovae with progenitor
masses $6\!-\!16\, M_{\odot}$ are ruled out as primary contributors to gas
with $\mbox{[$\alpha$/Fe]} \ga 0.4$ since their ejecta are not  
$\alpha$-enriched enough.

We assume that the extreme globular clusters considered here have similar
masses to Milky Way globular clusters. E.g., for a typical mass of
$\sim\!10^{5.5}M_{\odot}$, and measured metallicities of up to
[Z/H]~$\approx0$, the total {\it metal} content of the parent gas cloud
amounts to up to $\sim\!5600\, M_{\odot}$. One way of producing this
amount of metals is to assume a progenitor/enriching stellar population
composed of high-mass ($\ga20\, M_{\odot}$) stars only (i.e. a progenitor
stellar population with an IMF truncated at the lower end at $20\,
M_{\odot}$), combined with a sufficiently high star-formation rate in
order to enable the formation of such massive stars
\citep[see][]{goodwin05}. In any other scenario, the extreme \afe\ ratios
require extremely short enrichment timescales, i.e.~a very short time
between the SN explosion of massive stars and the formation of globular
clusters: of the order of the lifetime of a $20 M_{\odot}$ star,
corresponding to a few Myr.

Our empirical result that globular clusters with such extreme chemical
compositions mainly reside in early-type galaxies but seem absent in
spirals suggests a fundamental difference in enrichment timescales 
between some phase(s) of globular cluster system formation.

A bimodality in the \afe\ distribution, as hinted by the top right panel
of Fig.~\ref{ps:amah}, would imply a cessation of globular cluster
formation processes over a period short enough to be unresolved in age (on
the scale of Figure~\ref{ps:amac} $\sim10^9$ yr), but lasting long enough
to allow a significant chemical contribution from lower-mass stars
($\la\!20\, M_{\odot}$, i.e.~few $10^6$ yr).~Speculating whether such a
delay could occur in the course of massive galaxy evolution, we note that
it might be introduced by the reionization epoch at redshifts
$6\!<\!z\!\la\!20$, which for a few $10^{7}$ years may have halted, or at
least suppressed, the formation of massive star clusters and enabled the
generation of less $\alpha$-enriched globular clusters which seem to
dominate GCSs in lenticulars and spirals.

\subsection{Comparison with chemical evolution models}
We compare our observational result for globular clusters with
photochemical evolution models of elliptical galaxies by \cite{pipino04}.
In Figure~\ref{ps:amac} we plot the predictions for \afe\ ratios for two
galaxies with $10^{10}$ and $10^{12} M_{\odot}$ initial baryonic mass. The
direct comparison shows that extreme \afe\ ratios can only be achieved
within the first $\la0.2$ Gyr after the ignition of star formation
processes. Further, the models predict similar total metallicities as
those of the most metal-rich extremely $\alpha$-enhanced globular cluster
in our sample. Note that the absolute starburst ignition time is
arbitrarily set in the models; i.e., one can delay the starburst ignition
time by an arbitrary amount which would leave the predictions of \afe\ as
a function of time {\it after\/} the burst unchanged. The delay between
the bursts of $10^{12} M_{\odot}$ and $10^{10} M_{\odot}$, on the other
hand, {\it is\/} fixed in the models, and depends on the total gas mass in
a given halo. This implies that less-massive starbursts, occurring later
from material with a similar chemical composition as for more massive
starbursts, can produce globular clusters with similarly high \afe\
ratios, but with younger ages. In a hierarchical galaxy formation picture
these younger clusters are subsequently merged into more massive halos.

It is important to realize that the very high \afe\ ratios for old
globular clusters in elliptical galaxies are {\it not} mirrored in the
{\it average} \afe\ of the field stellar population as measured from the
diffuse light.~Using the same SSP models as in this study, \cite{thomas05}
recently found that early-type galaxies have lower luminosity-weighted
\afe\ values ($\langle$[$\alpha$/Fe]$\rangle_{\rm E}\approx0.25$) with
very few galaxies showing \afe~$\!>\!0.3$ dex, i.e., values comparable to
those of the low-\afe\ peak of globular clusters in ellipticals and most
clusters in lenticular and spiral galaxies. This suggests that the
formation of massive globular clusters with extremely high \afe\ ratios
predates the formation of the {\it majority of stars} in elliptical
galaxies.~However, we emphasize that our findings do {\it not} exclude the
presence of some extremely $\alpha$-enriched field stars in elliptical
galaxies.

Interestingly, this would be consistent with the popular scenario in which
stars are born in star clusters and gradually become part of the diffuse
light through evaporation by two-body relaxation and tidal stripping of
the star clusters \citep[e.g.,][]{lada03,fall05}.~Indeed, the timescale of
such disruption mechanisms is very long (more than a Hubble time) for
massive globular clusters of $\sim$\,10$^{5.5} M_{\odot}$ and shorter 
for less massive ones \citep[$\dot{M}/{M}
\propto M^{-1/2}$ for two-body relaxation; e.g.][]{fall01}, so that most
of the diffuse light in ellipticals would originate, in such a scenario,
from lower-mass clusters, most of which dissolved after a few Gyr or
sooner \citep[see also][]{goudfrooij04}.~This actually suggests that
lower-mass clusters in ellipticals should generally {\it not\/} show \afe\
$\ga$ 0.4, which would imply a later formation. These predictions 
should be testable in the near future.

\section{Summary}
We present \afe, metallicity, and age distributions of globular clusters
in elliptical, lenticular, and spiral galaxies, that were derived from
Lick line index measurement and identify a population of globular clusters
with extreme \afe\ ratios.~The main results of our study of extremely
$\alpha$-enriched globular clusters are: {\it i)} they are predominantly
found in early-type galaxies, {\it ii)} most of these highly
$\alpha$-enriched globular clusters are old ($t\!>\!8$ Gyr) and cover the
metallicity range $-1\!\la$ \zh\ $\la\!0$, {\it iii)} a comparison with
supernova yield models suggests that the progenitor gas clouds of these
globular clusters were predominantly enriched by massive stars ($\ga\!20
M_{\odot}$), {\it iv)} given the lower {\it average} \afe\ ratios of the
diffuse stellar population in early-type galaxies, our results suggest
that the extremely $\alpha$-enhanced globular clusters are members of the
very first generation of star clusters formed, and that their formation
epochs likely predate the formation of the {\it majority of field stars}
in giant early-type galaxies.

\acknowledgments

THP acknowledges financial support in form of an ESA Research
Fellowship.~We thank Antonio Pipino and Francesca Matteucci for providing
their models in electronic form and useful discussions.~This research has made use of the NASA/IPAC Extragalactic Database (NED) which is operated
by the Jet Propulsion Laboratory, California Institute of Technology,
under contract with the National Aeronautics and Space Administration.

\end{document}